\documentstyle[epsfig]{mn} %\documentstyle[referee,epsfig]{mn}
 \newcommand{\mnras}{MNRAS}
\newcommand{\apj}{ApJ} 

\title[] {Asymmetries in the inner regions of $\Lambda$CDM haloes}

\author[L. ~Gao, S.~D.~M. White] {Liang ~Gao$^{1,2}$ \thanks{Email:
    liang.gao@durham.ac.uk}, Simon D.~M.~White$^1$
\\ $^1$ Max--Planck--Institut f\"ur Astrophysik, D-85748, Garching, Germany
\\ $^2$ Institute for Computational Cosmology, Department of Physics,
Univsersity of Durham, South Road, Durham, DH1 3LE}
\begin{document} \label{firstpage} \maketitle

\begin{abstract}
Many galaxies display warps, lopsided images, asymmetric rotation
curves or other features which suggest that their immediate dynamical
environment is neither static nor in equilibrium. In Cold Dark Matter
(CDM) theories, such non-equilibrium features are expected in the
inner regions of many dark haloes as a result of recent hierarchical
growth. We used the excellent statistics provided by the very large
Millennium Simulation to study (i) how the distribution of position
and velocity asymmetries predicted for halo cores by the concordance
$\Lambda$CDM cosmogony depends on halo mass, and (ii) how much of the
dark matter in the inner core has been added at relatively recent
times. Asymmetries are typically larger in more massive haloes. Thus
$20\%$ of cluster halos have density centre separated from barycentre by
more than $20\%$ of the virial radius, while only $7\%$ of Milky Way halos
have such large asymmetries. About $40\%$ of all cluster halos have a
mean core velocity which differs from the barycentre velocity by more
than a quarter of the characteristic halo circular velocity, whereas
only $10\%$ of Milky Way halos have such large velocity offsets. About
$25\%$ of all cluster haloes have acquired more than a quarter of the
mass currently in their inner $10$kpc through mergers since $z=1$. The
corresponding percentage of Milky Way haloes is $15\%$. These numbers
seem quite compatible with the levels of asymmetry seen in the
observable regions of galaxies, but quantitative comparison requires
more detailed modelling of the observable components.
\end{abstract}

 \title{Asymmetries in the inner regions of $\Lambda$CDM haloes}

\section{Introduction}

Many spiral galaxies (including our own) display marked warps.  These are
typically of ``integral sign'' type with the outer part of the galactic disk
bending above its principal plane on one side of the system and bending below
it on the opposite side. These features appear to obey a number of
regularities, but the processes by which they are excited and maintained
remain uncertain (Toomre 1983; Briggs 1990; Garcia-Ruiz, Sancisi \& Kuijken
2002).  Another common asymmetry is lopsidedness -- the disk extends to
greater distances on one side of the centre than on the other -- the prototype
system being M101, the nearest giant Sc galaxy. Near-infrared imaging shows
that the asymmetry affects old disk stars as well as gas and star-forming
regions, and that amplitudes above 20\% are found in 30\% of normal nearby
spirals (Zaritsky \& Rix 1997; Bournaud et al 2005). Related irregularities
show up in kinematic maps of galaxies, where both optical and radio data
indicate that the major-axis ``rotation curves'' of galaxies can appear quite
different on opposite sides of the centre (Verheijen 1997; Swaters et al 1999;
Rubin, Waterman \& Kenney 1999).

Such apparent departures from symmetry are not surprising in the
standard Cold Dark Matter (CDM) cosmogony, where low mass objects
collapse first and then merge and accrete in a highly inhomogeneous
fashion to form larger and larger systems. As each object collapses
and grows, it experiences violent relaxation leading to a temporary
quasi-equilibrium state with near-universal structure (Navarro, Frenk
\& White 1997). Infalling previously virialized systems typically
survive as self-bound substructures for a number of orbits within
their new host before finally dissolving.  In high-resolution
simulations the fraction of the mass of a dark matter halo in such
substructures ranges from 1\% to 20\%, with most of it lying in
the few most massive objects (e.g. Gao et al. 2004a; Diemand, Moore \&
Stadel 2004).  As a result of such ongoing and inhomogeneous accretion
processes the dynamical structure of CDM halos deviates significantly
from true equilibrium.  These deviations can be thought of as
excitations of the underlying equilibrium model which may be related
to warps or lopsidedness in the central galaxy (Debattista \& Sellwood
1999; Jog 2002).

In this paper, we use a very large cosmological simulation to study
asymmetries of dark matter haloes which may be related to such distortions
of their central galaxies. The simulation is the so-called
``Millennium Simulation'' carried out by the Virgo Consortium (Springel et
al. 2005).  This simulation adopted concordance values for the parameters of a
flat $\Lambda$CDM cosmological model, $\Omega_{\rm dm}=0.205$, $\Omega_{\rm
b}=0.045$ for the current densities in Cold Dark Matter and baryons, $h=0.73$
for the present dimensionless value of the Hubble constant, $\sigma_8=0.9$ for
the {\it rms} linear mass fluctuation in a sphere of radius $8 h^{-1}$Mpc
extrapolated to $z=0$, and $n=1$ for the slope of the primordial fluctuation
spectrum. The simulation followed $2160^3$ dark matter particles from $z=127$
to the present-day within a cubic region $500 h^{-1}$Mpc on a side. The
individual particle mass is thus $8.6\times 10^{8}h^{-1} {\rm M_\odot}$, and
the gravitational force had a Plummer-equivalent comoving softening of
$5h^{-1}$kpc.  The {\small TREE-PM} N-body code {\small GADGET2}(Springel
2005) was used to carry out the simulation.

The excellent statistics provided by the Millennium Simulation allow us to
study (i) the distributions of position and velocity asymmetry predicted for
halo cores by the concordance $\Lambda$CDM cosmogony, (ii) the dependence of
these distributions on halo mass, and (iii) the amount of material added to
the inner core at relatively recent times. Our paper is structured as
follows. In Section 2, we describe how halos are defined in the Millennium
Simulation and how we measure offsets in position and velocity for their inner
cores relative to their main bodies. We present results for these spatial and
velocity asymmetries as a function of halo mass in Section~3. In Section~4, we
study the accretion of material into the inner core. Finally we discuss our
results and set out our conclusions in Section~5.

\section {Methods}
\subsection{Halo and subhalo catalogues in the Millennium Simulation}
Nonlinear objects (``halos'') can be identified in numerical simulations by a
variety of methods. Two of the most common are the Friends-of-Friends scheme
({\small FOF} Davis et al. 1985) which links together all particle pairs
closer than some chosen limit and defines halos as disjoint sets of mutually
linked particles, and the spherical overdensity scheme ({\small SO} Lacey \&
Cole 1996) which defines halo centres as local potential minima or density
maxima and halo boundaries as the largest spheres surrounding these centres
for which the mean enclosed density exceeds a chosen threshold; halos whose
centre lies inside a more massive halo are then discarded. Both methods have
both pros and cons. {\small FOF} selection does not impose any fixed shape or
symmetry on the halos and does not require any {\it a priori} choice of centre
or any pruning of the halo catalogue.  On the other hand, its halos are often
made up of distinct clumps with well-separated centres joined by relatively
low-density filaments. The {\small SO} scheme does not have this problem but
it chooses an {\it a priori} centre and requires the halo boundary to be
spherically symmetric about this centre.

The halos we analyse in the Millennium Simulation were identified
in a more complex way, described in detail by Springel et al.
(2005). Particle groups are defined with a {\small FOF} linking
length $0.2$ times the mean particle separation. Using the
algorithm {\small SUBFIND} (Springel et al. 2001) each {\small
FOF} group is then separated into a set of disjoint, locally
over-dense, self-bound substructures and a (typically small)
number of unbound particles.  Most groups then consist of a single
dominant (sub)structure, which can be identified as the main halo,
a set of much smaller substructures, and some diffuse unbound
material; groups where the {\small FOF} algorithm joined distinct
objects are broken into their constituent parts.  As part of this
procedure, binding energies are computed and stored for all
particles within each (sub)structure. We use these below.  For
typical halos about $90 \%$ of the mass is in the main halo. The
fraction of mass in substructure correlates quite strongly with
halo formation time (Gao et al. 2004a; Zentner et al. 2005; Shaw
et al. 2006). Haloes that formed earlier tend to contain less
substructure. This is easily understood as a consequence of the
dynamical disruption of accreted objects (Gao et al. 2004a; Taylor \&
Babul 2004; Zentner et al. 2005; Van den Bosch et al. 2005a).

\subsection{Position offsets for halo cores}

As an operational definition of position of the ``core'' of each halo
we use the average position of the $100$ most bound members of its
main subhalo. This is motivated by several considerations. As we will
see below, the average distance of these 100 particles from their
barycentre is about $10 {h^{-1}\rm kpc}$, independent of halo mass.
This choice thus defines a core of similar size to the regions for
which distortions are measured in real galaxies but comfortably larger
than the gravitational softening scale of the simulation ($5
h^{-1}$kpc). In addition, 100 particles is enough that noise due to
discreteness effects is negligible. Indeed, in almost all halos the
core position so defined is very close to that of the single particle
with the lowest gravitational potential or with the highest local
density.

We consider two different definitions of the position of a halo as a whole.
The first is the barycentre of all the particles assigned to its main subhalo
by {\small SUBFIND}. This excludes {\small FOF} group members which are either
part of a smaller substructure or are unbound to any substructure. This
definition does not make any {\it a priori} assumption about the symmetry of
the halo, but it excludes material which is most naturally associated with
another ``object''.  It may be the natural definition to use if one wishes to
compare with analyses which model the distortion of a galaxy (for example, the
warping of the Milky Way due to tidal interaction with the Magellanic Clouds)
as excitations of a regular system driven by external gravitational forcing.
We dub this barycentre $r_{\rm main}$, so the position offset between
the halo and its core $\Delta r_{\rm main}$ can be written as
\begin{equation}
\Delta r_{\rm main}=|\vec{r}_{\rm core}-\vec{r}_{\rm main}| ,
\end{equation}
where $\vec{r}_{\rm core}$ is the barycentre of the $100$ most
bound particles in the main subhalo and $\vec{r}_{\rm main}$ is
that of the main subhalo as a whole.

For comparison, we will also show some results for offsets where the centre of
a halo is defined as the barycentre of the corresponding {\small SO}
group. For this purpose we take the group centre to be the
{\small FOF} particle with the greatest potential energy and we define the
halo as all particles (including substructures and unbound particles) within
the largest sphere for which the mean enclosed overdensity is at least 200
times the critical value. With this definition the core offset is
\begin{equation}
\Delta r_{\rm so}=|\vec{r}_{\rm core}-\vec{r}_{\rm so}| ,
\end{equation}
where $\vec{r}_{\rm so}$ is the barycentre of all members of the {\small SO}
halo. Clearly since the {\small SO} halo is bounded by a sphere which is
effectively centred on $\vec{r}_{\rm core}$, we can expect typical offsets to
be smaller in this case than with our preferred definition.

\subsection{Velocity offsets for halo cores}
We define mean velocities for halos as a whole in direct analogy to the mean
positions defined above by averaging either over all particles of the main
subhalo ($\vec{V}_{\rm main}$) or over all particles of the {\small SO} halo
($\vec{V}_{\rm so}$).  Note that again the first definition excludes
substructures but the second does not. Measuring a velocity offset for the
core is more difficult than measuring a position offset because of the
``noise'' introduced by the large random motions of particles in the inner
halo. We rank particles by their distance from the centre of the core (taken
as $\vec{r}_{\rm core}$) and we estimate the square of the velocity offset for
the $N$ innermost particles as
\begin{equation}
  \Delta V^2(N)=|\vec{V}_{\rm core}-\vec{V}_{\rm
    bulk}|^2-\frac{1}{N}\sigma^2_{\rm core}, \\
\end{equation}
where $\vec{V}_{\rm bulk}$ is either $\vec{V}_{\rm main}$ or $\vec{V}_{\rm
so}$, and $\vec{V}_{\rm core}$ and $\sigma_{\rm core}$ are defined by
\begin{equation}
  \vec{V}_{\rm core} = N^{-1}\sum_{i=1}^N\vec{v}_{\rm i}
\end{equation}
and
\begin{equation}
  \sigma^2_{\rm core} = (N-1)^{-1}\sum_{i=1}^N|\vec{v}_{\rm i}-\vec{V}_{\rm
      core}|^2.
\end{equation}
With these definitions we expect our estimator of velocity offset (squared) to
be unbiased but it will give negative values for some halos. Choosing large
$N$ will reduce the noise but will result in overly large ``cores'' for the
lower mass halos. We investigate the appropriate compromise below.

\section{Results}
\subsection{Position and velocity asymmetries}

We now examine how the above position and velocity offsets are distributed for
large samples of halos drawn from the Millennium Simulation. It is clearly of
interest to understand how such distributions depend on halo mass. Naively,
lower mass haloes were assembled earlier than more massive ones, so it seems
natural that they should typically be more relaxed and have smaller
asymmetries both in position and in velocity.

We select four sets of halos randomly from the Millennium Simulation in four
different mass ranges. There are $1636$ ``Milky Way'' haloes with $M_{200}$ in
the range $[2.0,4.0] \times 10^{12}h^{-1}{\rm M_\odot}$; there are $640$
``poor group'' halos in the mass range $[0.7,2.0] \times 10^{13}h^{-1}{\rm
M_\odot}$; there are $280$ ``rich group'' halos in the mass range $[0.7,2.0]
\times 10^{14}h^{-1}{\rm M_\odot}$; finally, there are $227$ ``cluster'' halos
with masses greater than $2 \times 10^{14}h^{-1}{\rm M_\odot}$.

\subsubsection {Position asymmetries}

\begin{figure*} \hspace{-0.5cm}
\resizebox{9cm}{!}{\includegraphics{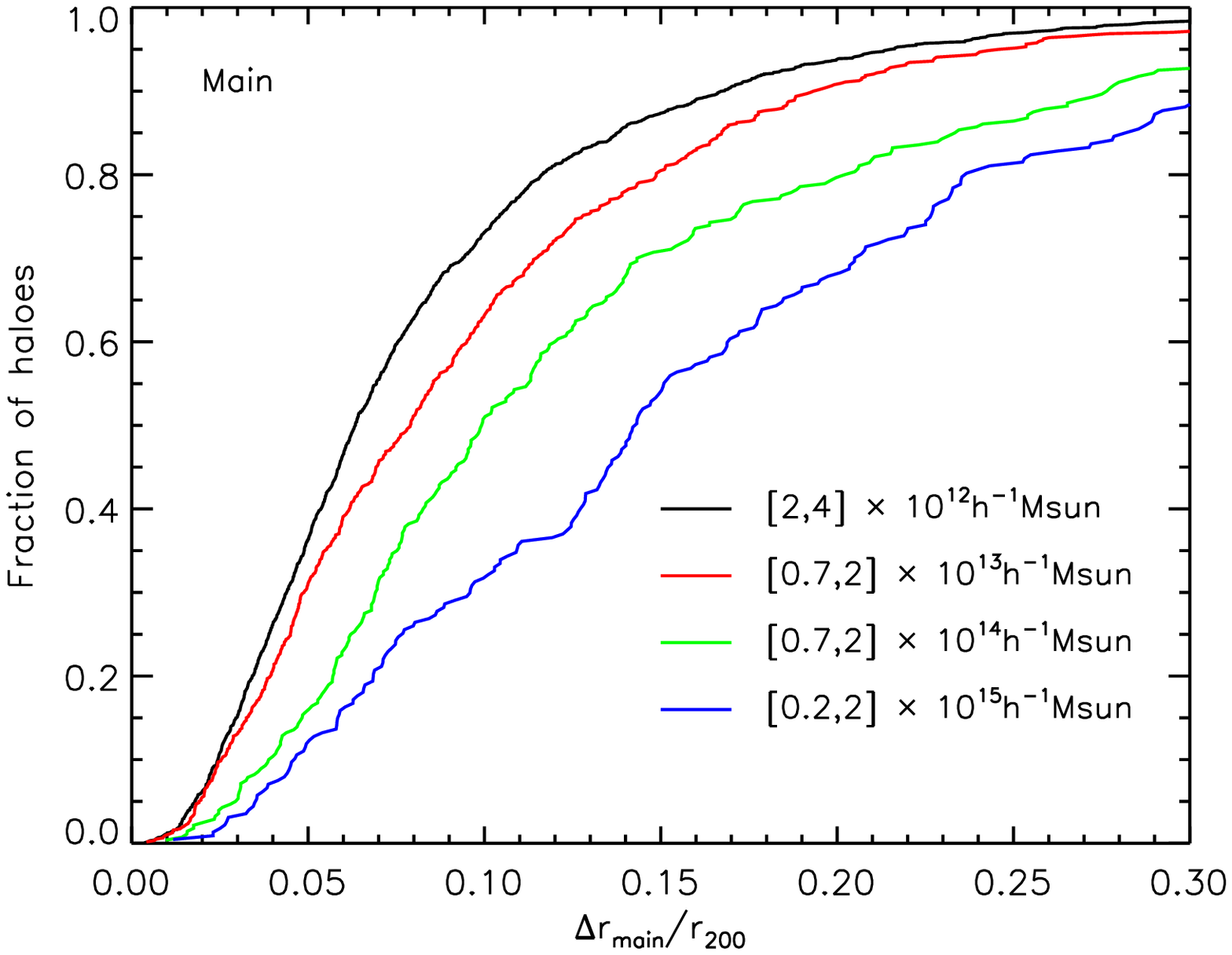}}
\hspace{-0.1cm}\resizebox{9cm}{!}{\includegraphics{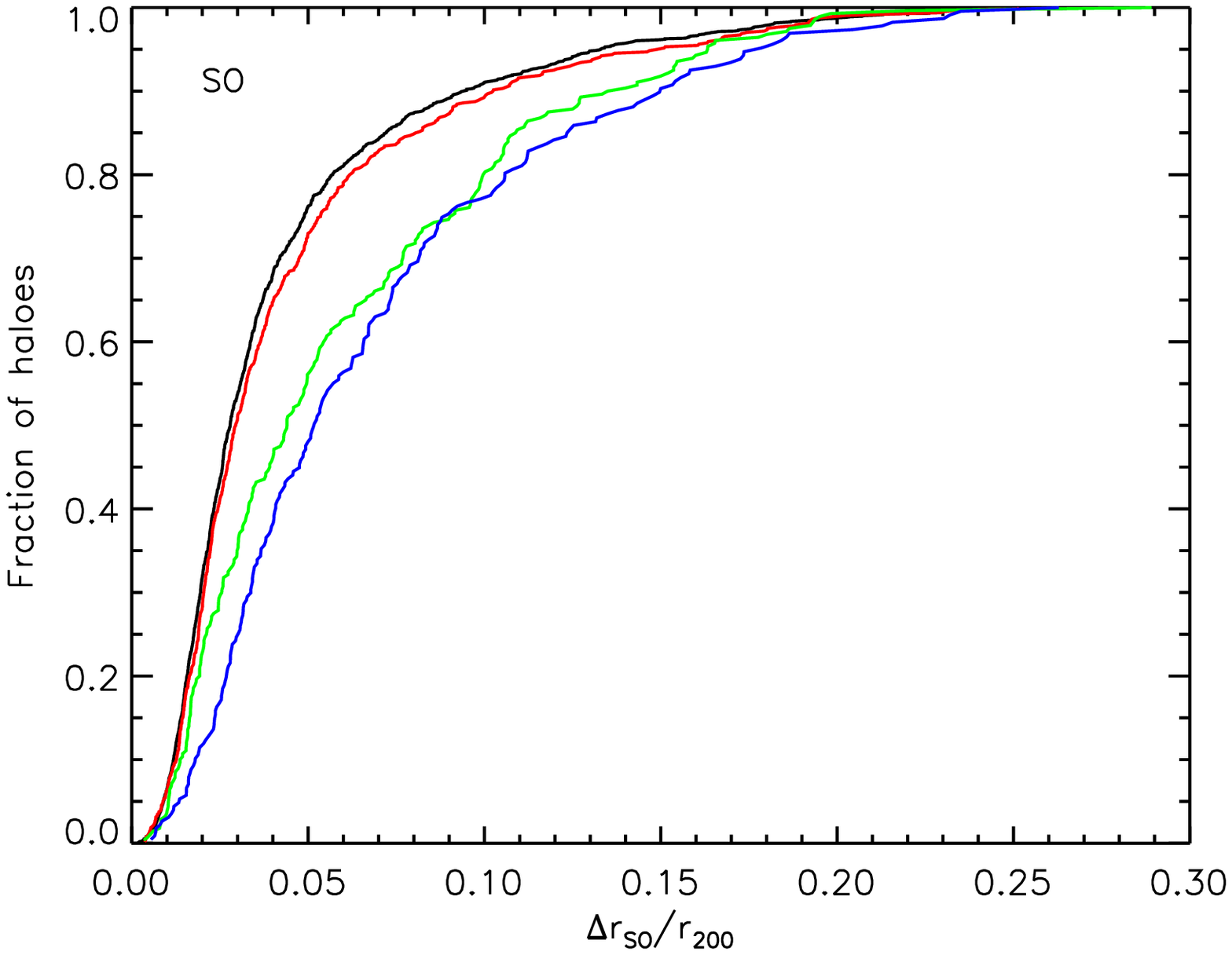}}\\
\caption{Cumulative distribution of halos as a function of position offset
  between the core and the bulk of the halo. In the left panel the halo
  position is defined as the barycentre of the main subhalo of the {\small
  FOF} group, while in the right panel it is defined as the barycentre of the
  {\small SO} group. Different line types in each panel refer to halos in
  different mass ranges as indicated. For all halos the position offset is
  expressed as a fraction of $r_{200}$, the radius of the corresponding
  {\small SO} group.}
\label{fig:fig1}
\end{figure*}

Cumulative distributions of the spatial offset between the barycentre of the
core and that of the halo as a whole are plotted in Figure~\ref{fig:fig1} for
halos in our four disjoint mass ranges. In the left panel the centre of each
halo is taken to be $\vec{r}_{\rm main}$, while in the right panel it is taken
to be $\vec{r}_{\rm so}$. In order to facilitate comparison of the different
mass ranges, the offset for each halo is expressed as a fraction of $r_{200}$,
the radius of the corresponding {\small SO} group. The offsets are
substantially smaller when we use $\vec{r}_{\rm so}$ to define the halo
centre, so the symmetry imposed artificially by the spherical boundary assumed
for the {\small SO} halos clearly affects the results much more strongly than
the omission of substructure when calculating $\vec{r}_{\rm main}$. For the
reasons discussed above, we consider offsets based on $\vec{r}_{\rm main}$ to
be the appropriate indicator of the kind of asymmetry which could drive galaxy
distortions, so we concentrate on results in the left panel of
Figure~\ref{fig:fig1} for the rest of this subsection.

It is clear that more massive haloes tend to have larger asymmetries. Thus the
cores of 20\% of cluster halos are offset from the barycentre of the main
subhalo by more than 20\% of $r_{200}$ ($\sim 200 h^{-1}$kpc), while only a
few percent of Milky Way haloes have such large asymmetries; the typical
offset for these lower mass halos is about 6\% of $r_{200}$ ($\sim 10
h^{-1}$kpc).  This mass dependence presumably reflects the fact that massive
halos typically assemble at later times and so are farther from equilibrium
today.  We come back to this issue below in our discussion section. Notice
that even for the Milky Way halos, the typical offsets are as large as the
visible size of the galaxy. In cluster halos they are easily large
enough to be measured reliably from X-ray images or lensing maps.

\begin{figure*}
\resizebox{8cm}{!}{\includegraphics{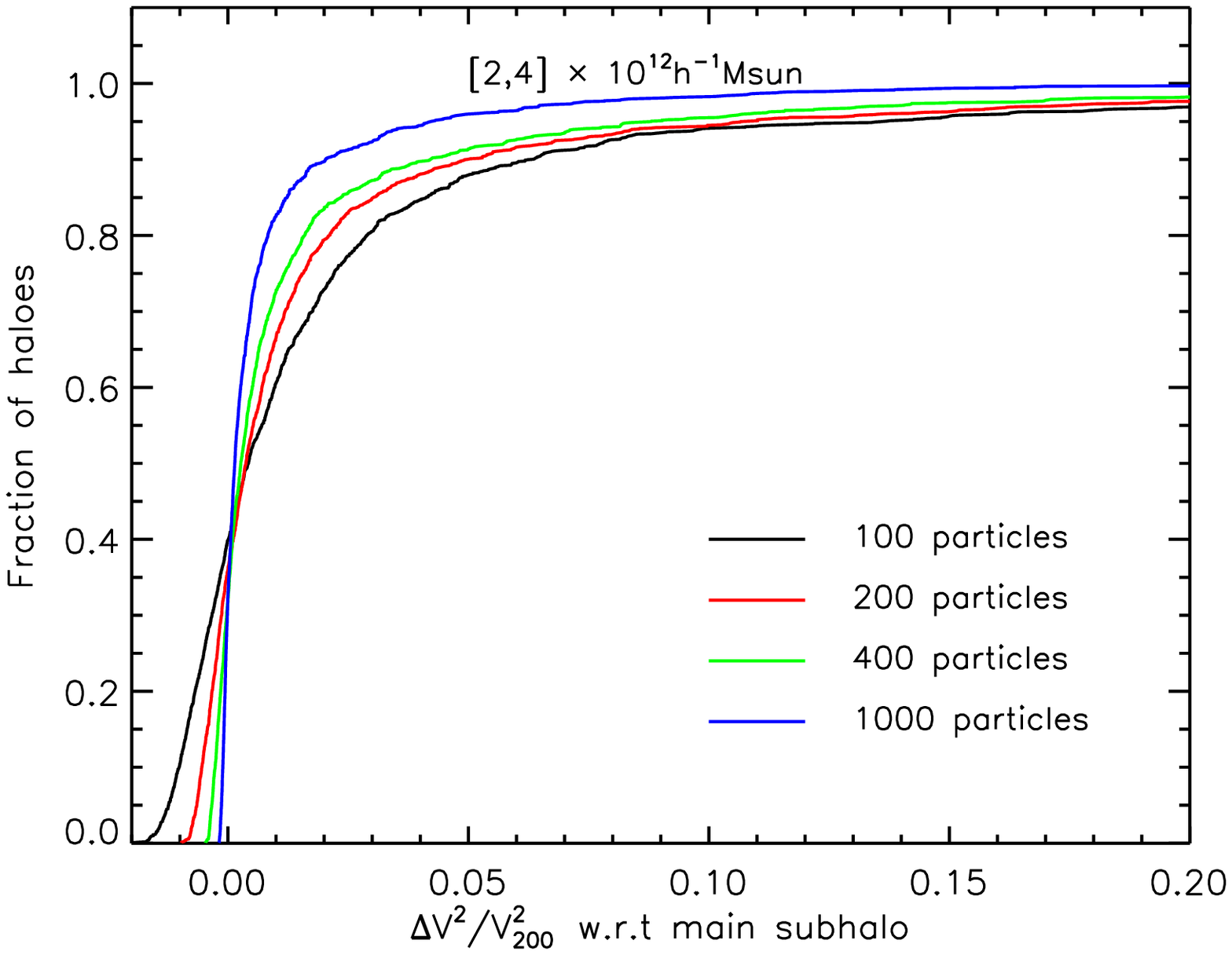}}
\hspace{0.13cm}\resizebox{8cm}{!}{\includegraphics{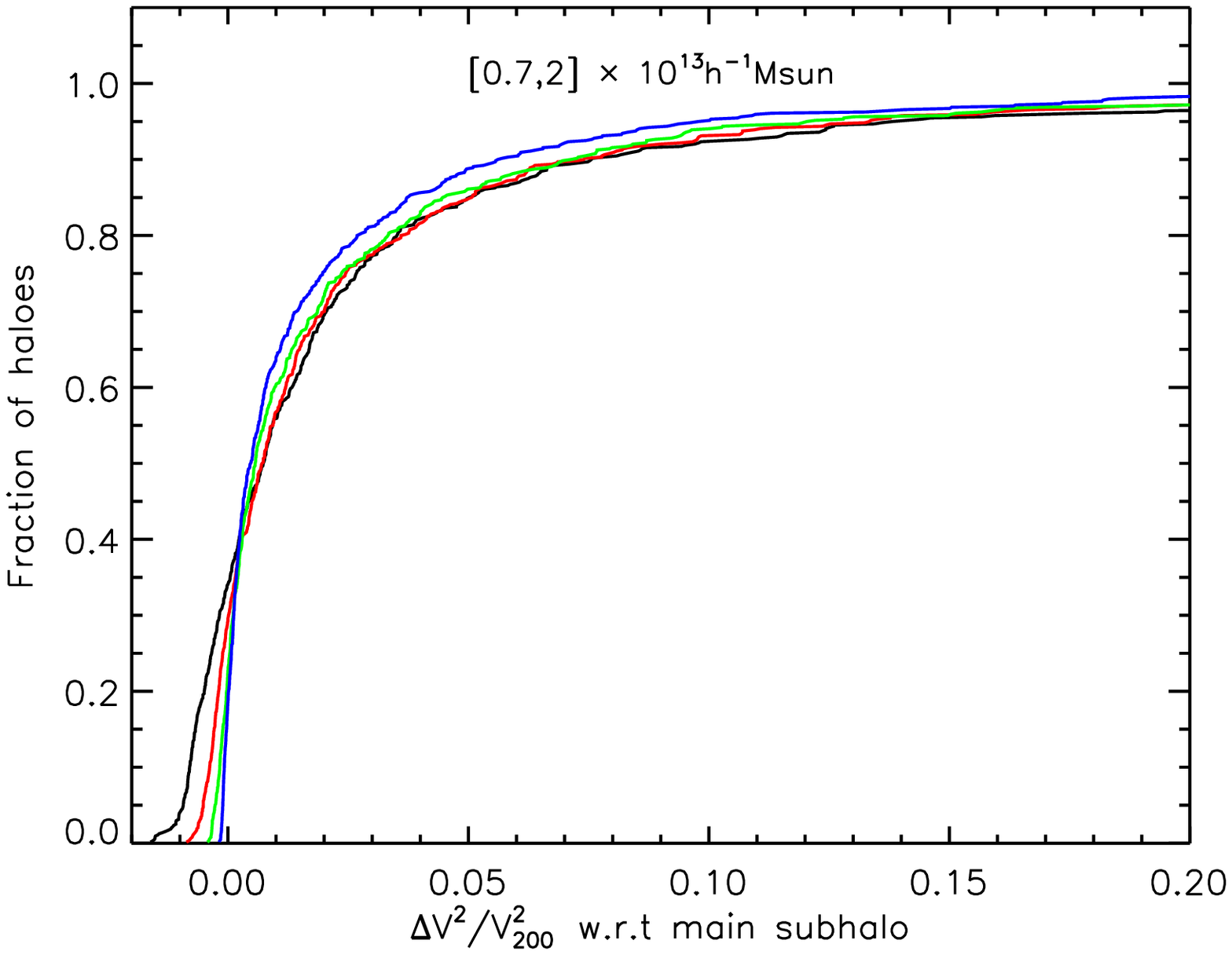}}
\resizebox{8cm}{!}{\includegraphics{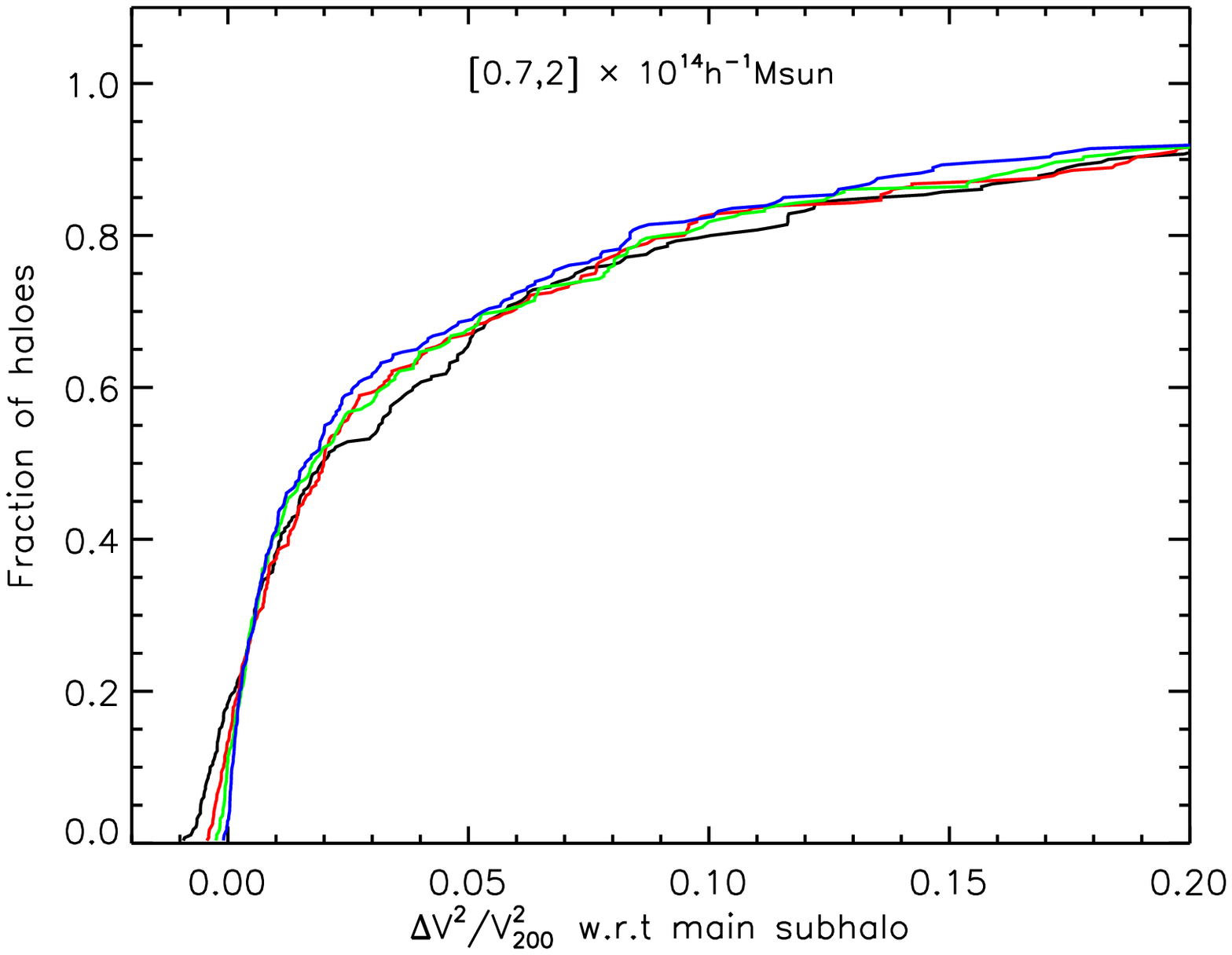}}
\hspace{0.13cm}\resizebox{8cm}{!}{\includegraphics{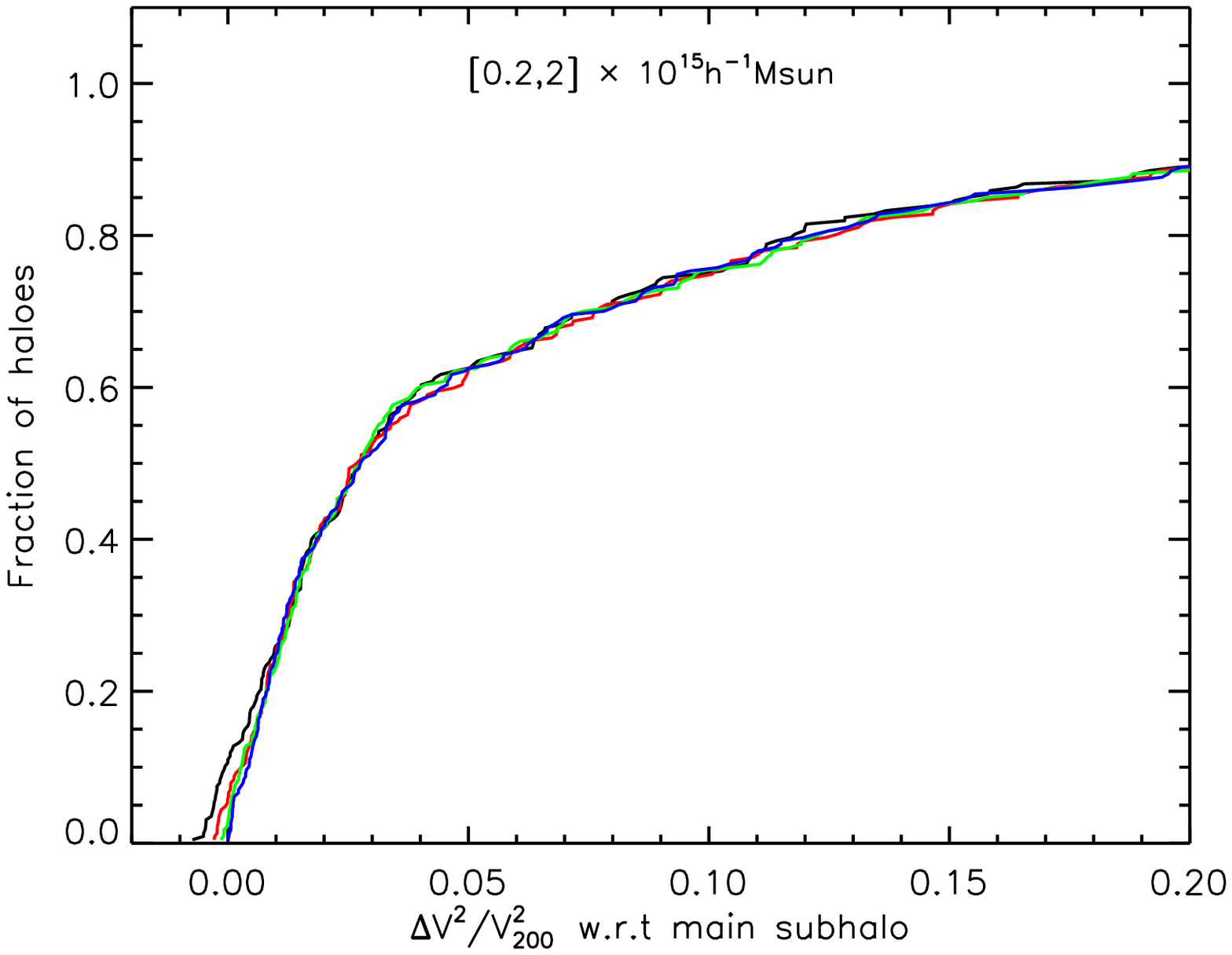}}

\caption{Cumulative distributions of the square of the velocity
offset between
  the core and the bulk of the main subhalo. The four panels refer to four
  different ranges of halo mass as noted.  In each panel curves show the
  distributions found for four different values of $N$, the number of
  particles averaged in determining the velocity of the core. (See the labels
  for the colour-coding.) Note that the values of $\Delta V^2$ have been
  corrected in the mean for noise in the core velocity measurement and so can
  be negative.  They have divided by $V_{200}^2$ to facilitate comparison of
  the different mass ranges.}
\label{fig:fig2}
\end{figure*}

\begin{figure*} \resizebox{8cm}{!}{\includegraphics{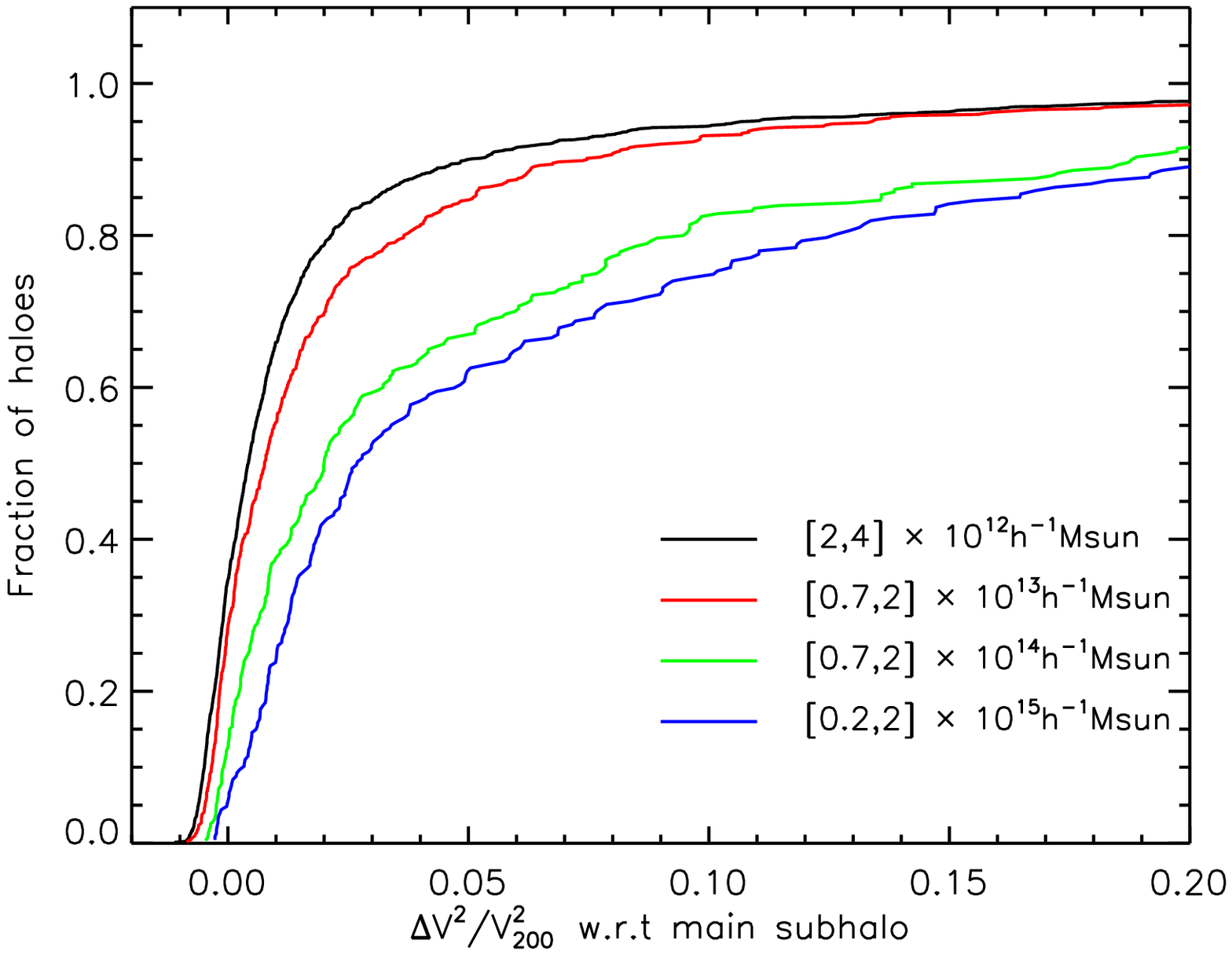}}
\hspace{0.13cm}\resizebox{8cm}{!}{\includegraphics{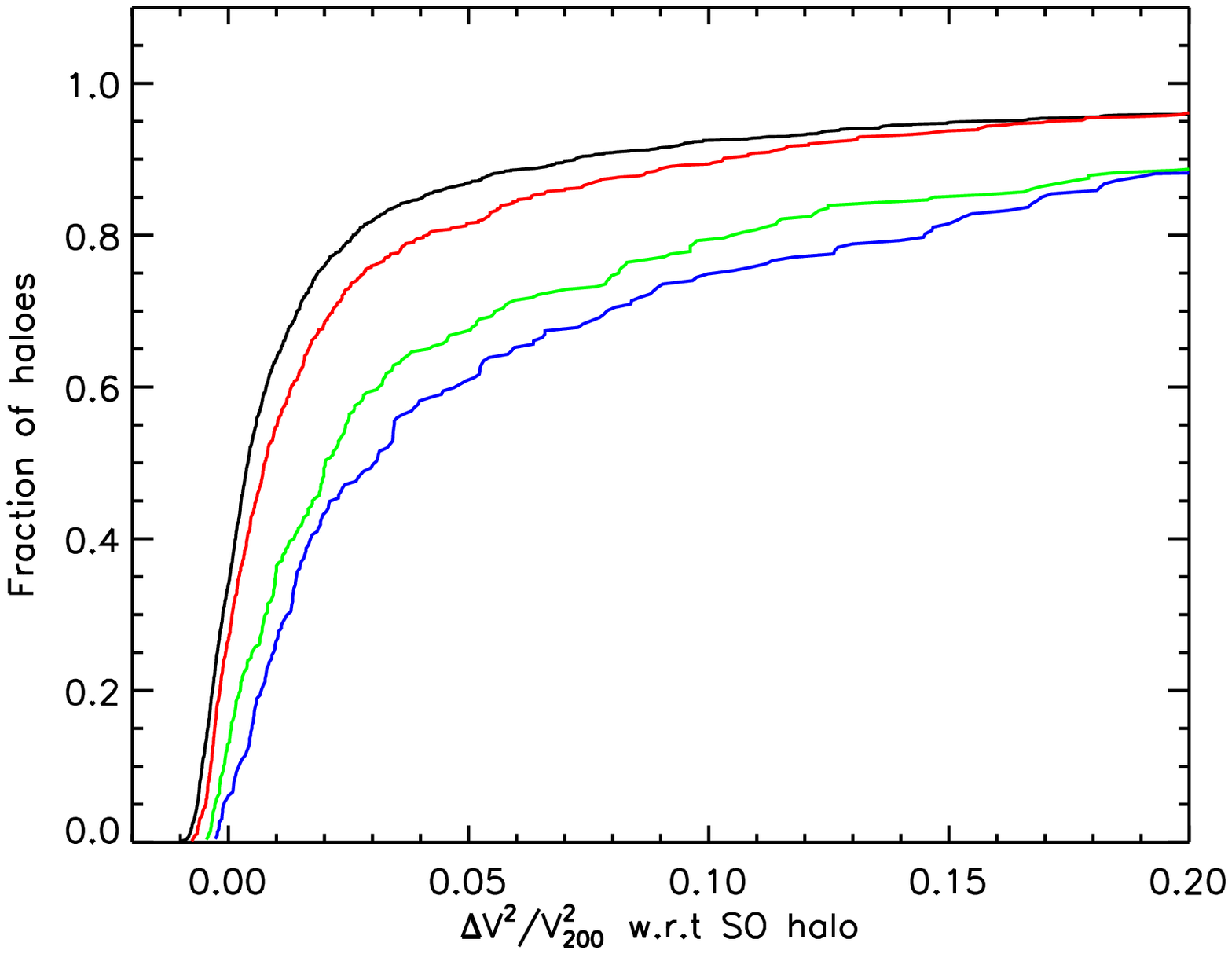}}
\caption{(left) Cumulative distributions of the square of the core
velocity
  offset. The left panel references the offset to the barycentric velocity of
  the main subhalo of each halo, and compares the $N=200$ curves from the four
  panels of Fig.~2. The right panel gives identical results except that the
  offset is now referenced to the barycentric motion of the {\small SO} halo.
  The offset for each halo has been normalised to $V_{200}$ to facilitate
  comparison. Labels indicate the mass range associated to each
  curve. Distributions with the two definitions of mean halo velocity are
  very similar.}
\label{fig:fig3}
\end{figure*}

\begin{figure*}%\Hspace{-0.5cm}
\resizebox{8cm}{!}{\includegraphics{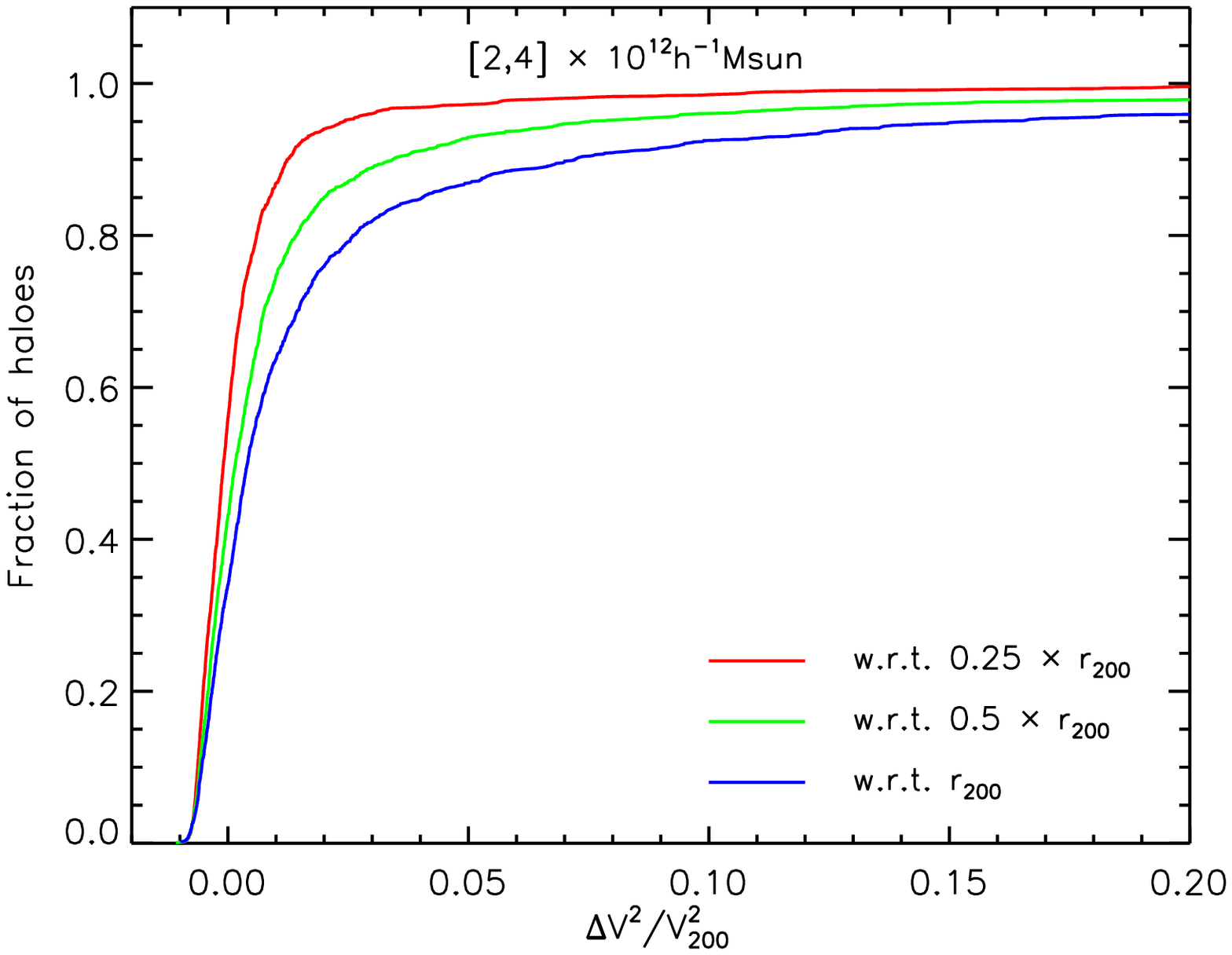}}%
\hspace{-0.1cm}\resizebox{8cm}{!}{\includegraphics{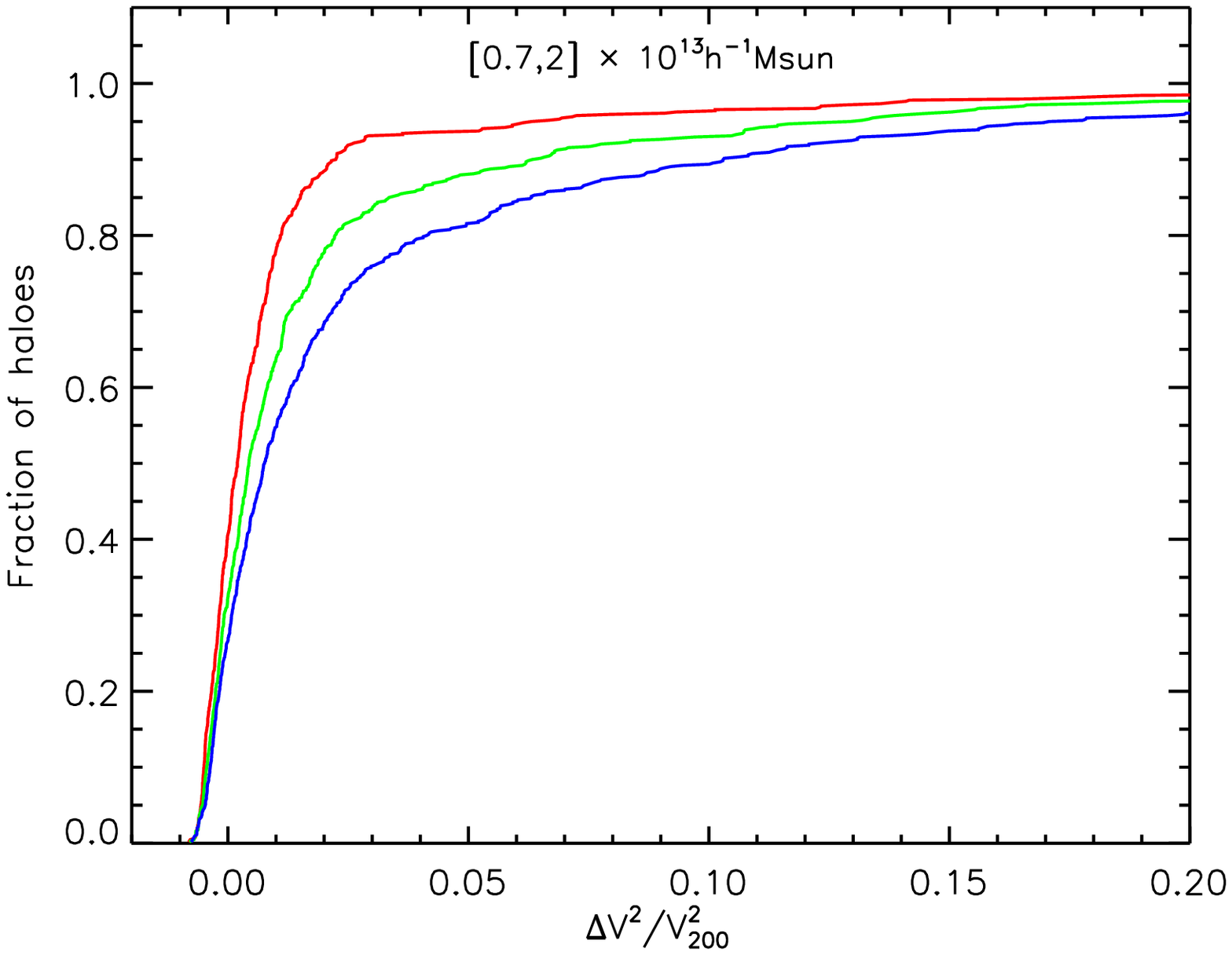}}\\%
\resizebox{8cm}{!}{\includegraphics{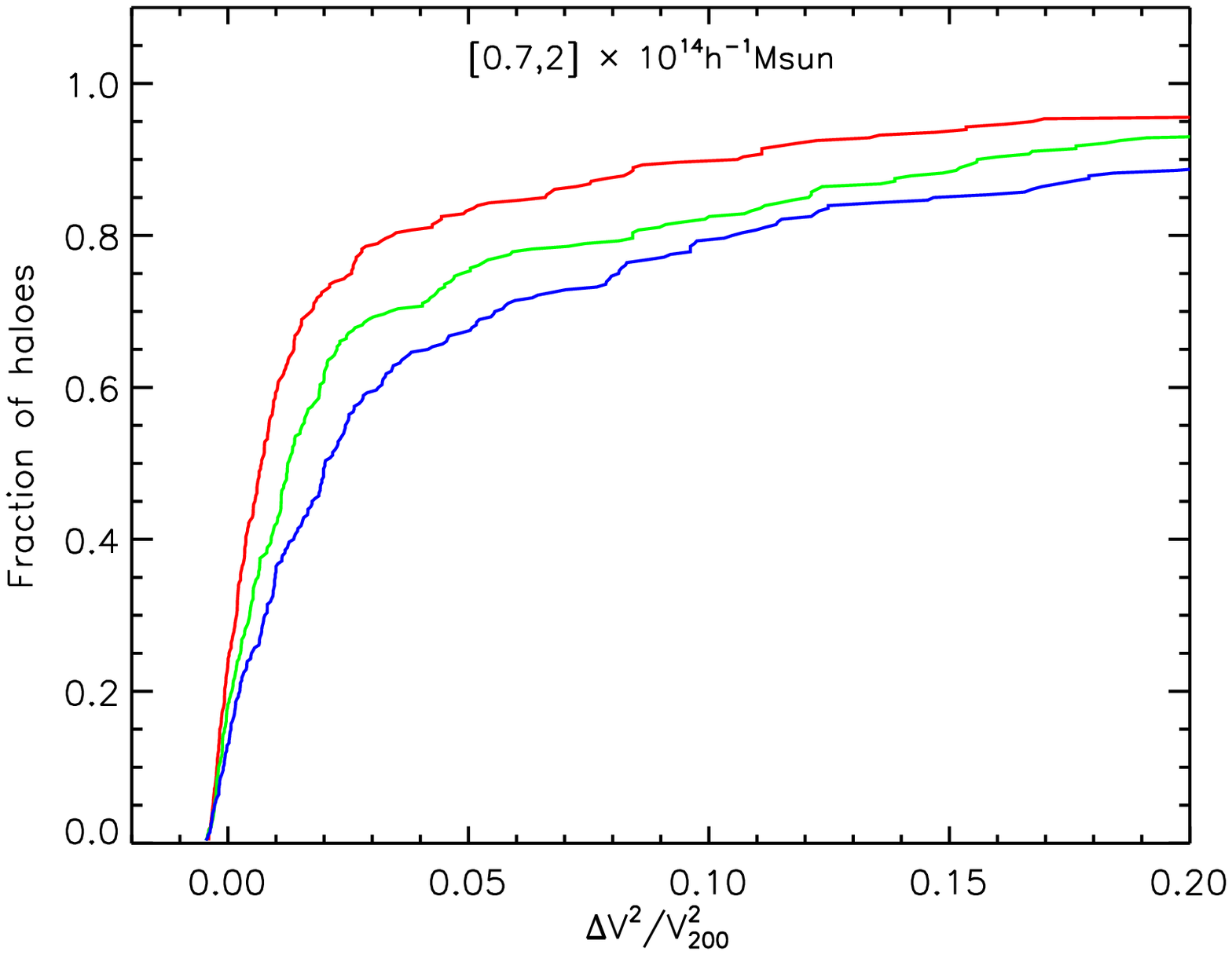}}%
\hspace{0.13cm}\resizebox{8cm}{!}{\includegraphics{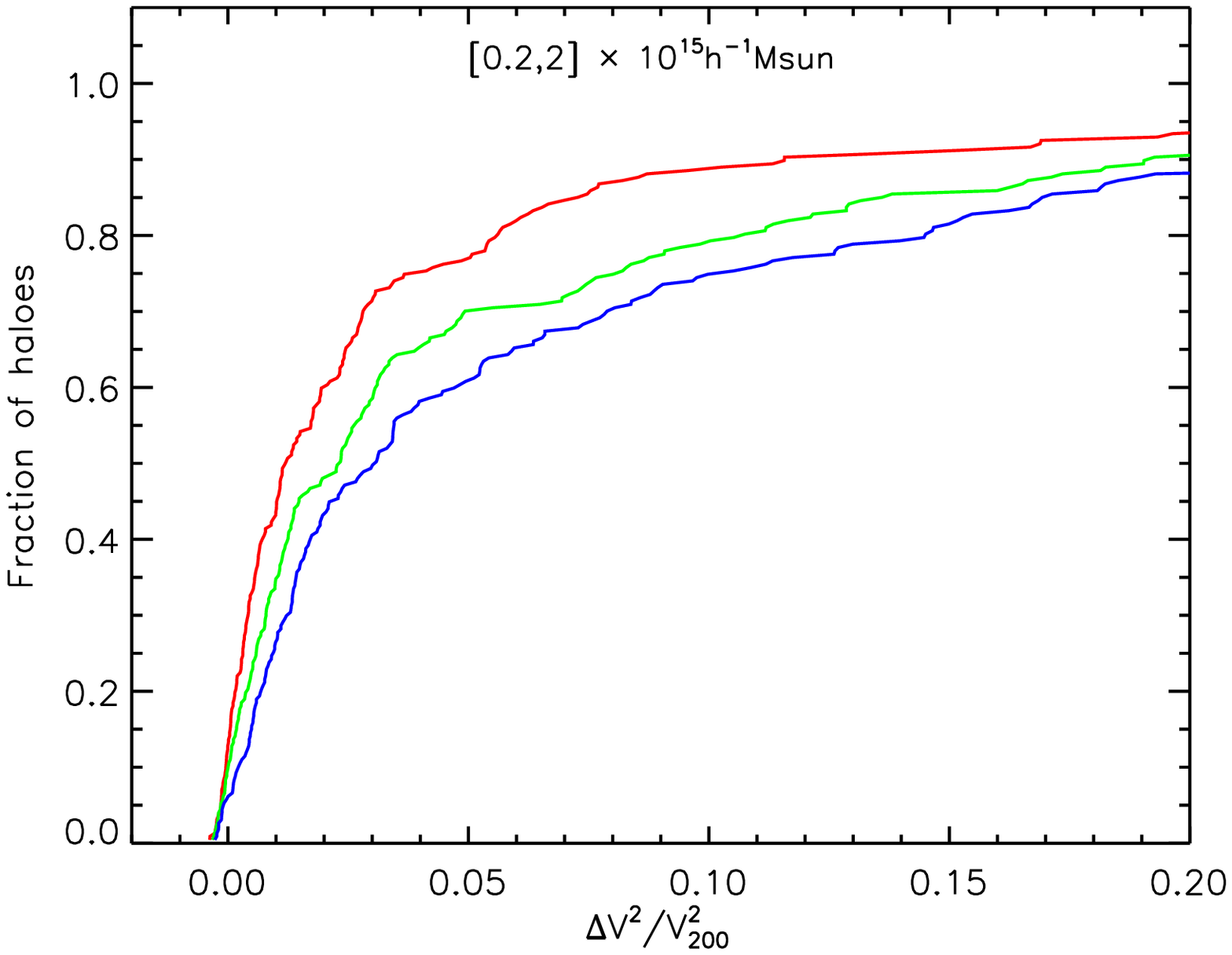}}
\caption{Cumulative distributions of the square of the velocity
offset of the
  core with respect to various fractions of the surrounding halo. Each panel
  refers to halos in one of our four mass ranges, as labelled. The core
  velocity is estimated using $N=200$ in all cases. The three curves in each
  panel show the distribution of the offset with respect to different
  fractions of the surrounding halo: all material within $r_{200}$
  (blue curves, repeating curves from the right panel of Fig.~3), all
  material within $0.5 r_{200}$ (green curves), and all material within
  $0.25 r_{200}$ (red curves).}
\label{fig:fig4}
\end{figure*}

\subsubsection{Velocity asymmetries}
We begin our study of velocity asymmetries by evaluating the
effects of using different values for $N$, the number of particles
used to define the mean velocity of the core in equations (3) to
(5). Large values of $N$ result in less measurement noise but an
overly large effective core for low-mass halos. Some compromise is
thus required for these systems.

In Fig.~\ref{fig:fig2} we show cumulative distributions for our
estimates of the square of the velocity offset between the core and
the bulk of the main subhalo in each of our objects (equation (3) with
 $\vec{V}_{\rm bulk}$ taken to be $\vec{V}_{\rm main}$). The four panels
refer to our four different halo mass ranges and the four curves in each
panel refer to different values of $N$. Here and below we divide the estimate
for each halo by $V_{200}^2 =   G M_{200}/r_{200}$ in order to make it
easier to compare results for the different mass ranges.

In the two bottom panels and for the three lowest $N$ values in the upper
right panel, the curves coincide within the noise for all but the smallest
velocity offsets.  This shows that for these $N$ the central region for which
the core velocity is estimated is small enough to be considered to move as a
unit. In the upper left panel and for the $N=1000$ curve in the upper right
panel a trend towards less extreme offsets for larger $N$ is visible. This is
because these halos have small enough masses ($\sim 3000$ particles on average
in the upper left panel) that increasing $N$ washes out a significant part of
the core motion. On the other hand, differences between the curves at small
velocity offset clearly show the effects of small-$N$ noise in our estimates
of core velocity. These are significant for $N=100$ but appear acceptably
small for $N\ge 200$, at least as judged from the curves for higher mass halos
which appear converged at large velocity offset. In the following we
adopt $N=200$ as the best compromise between these competing effects.

In the left panel of Fig.~\ref{fig:fig3}, we replot the $N=200$ curves of
Fig.~\ref{fig:fig2} on top of each other for easier comparison. There is a
clear systematic trend for more massive halos to have larger velocity
asymmetries, in direct analogy to the trend found above for position
asymmetries.  More than a quarter of all cluster halos have core velocities
which differ from the mean halo value by at least 30\% of $V_{200}$ (i.e. by
velocities greater than about 300 km/s), whereas only a few percent of Milky
Way halos have such a large offset. The typical offset for low mass halos is
small and only about 15\% of them have offsets exceeding $0.2 V_{200}$ (i.e.
greater than about 40 km/s). The right panel of Fig.~\ref{fig:fig3} shows
identical curves, except that the offset is now calculated with respect to the
barycentric motion of the {\small SO} halo. The resulting distributions are
almost indistinguishable from those in the left panel, showing that effects
due to substructures and to the definition of the halo boundary are too small
to be significant for these statistics.

An interesting question is whether the relative motions we measure are due to
non-equilibrium effects in the outer part of the halos or whether they also
reflect significant motions of the core with respect to intermediate halo
regions. Presumably motions of the latter type are more likely to relate to
observable galaxy distortions such as warps or lopsidedness.  In
Fig.~\ref{fig:fig4}, we address this question by measuring velocity offsets of
the core relative to different regions of the halo. The four panels here refer
to halos in each of our four mass ranges. The three curves in each panel give
the cumulative offset distributions for core velocities calculated relative to
all particles within $r_{200}$, relative to all particles within $0.5
r_{200}$, and relative to all particles within $0.25 r_{200}$. 

As expected, typical offsets go down in all cases as the size of the
reference region shrinks. For the lowest mass halos the reduction is a
factor of 2 to 3 in $\Delta V^2$ from the full {\small SO} halo to its
innermost 25\% (corresponding to a region about $50 h^{-1}$kpc in
radius surrounding the galaxy). Reductions are by somewhat smaller
factors for more massive halos. Nevertheless, quite substantial
motions are detected even for the smallest regions, so a significant
fraction of the core motion is typically relative to the immediately
surrounding halo.  A similar conclusion can be drawn from the upper
left panel of Fig.~2 which shows that the tail of large measured
motions shrinks when the effective size of the ``core'' is increased. 

\begin{figure*}
\hspace{8cm}\resizebox{16cm}{!}{\includegraphics{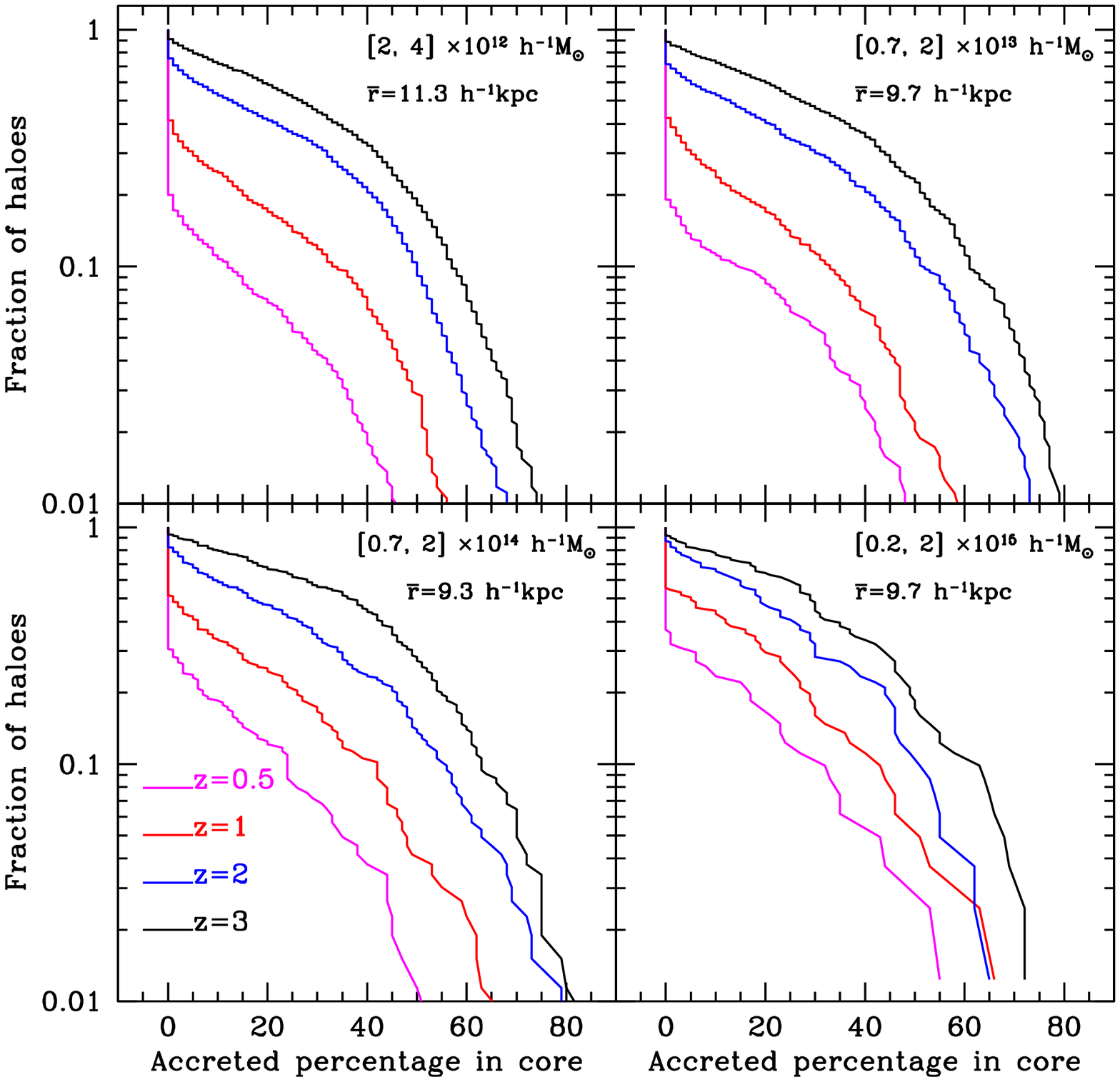}}
\caption{Cumulative distributions of the fraction of the $z=0$
core mass added
  by merger-driven accretion since a series of earlier redshifts. Each panel
  refers to halos in one of our four mass ranges and shows curves for
  accretion since the four redshifts indicated by the colour code. The $z=0$
  core of each halo is here defined to be the 100 most bound particles of its
  main subhalo. The accreted fraction is then the percentage of these
  particles which are more than $100$kpc (physical) from the centre of
  their main concentration at each earlier redshift. Labels in each panel give
  the mass range of the halos plotted and the mean distance of their ``core''
  particles from halo centre at $z=0$.}
\label{fig:fig5}
\end{figure*}
\subsection{Late accretion by cores}

So far we have addressed non-equilibrium excitations of the inner regions of
galaxy and cluster halos by looking directly for the position and velocity
asymmetries which they may produce.  In this subsection, we approach the issue
from a different angle by studying the rate at which material is added to the
core regions by the merger/accretion events which typically drive such
excitations.

In galaxies this process is related to the build up of the stellar halo
through accretion and disruption events like that currently involving the
Sagittarius dwarf galaxy (Ibata et al. 2001). In rich clusters it is related
to the formation of the central galaxy by cannibalism of other cluster members
(Ostriker \& Tremaine 1975; White 1976; Dubinski 1998).  Using a set of high
resolution resimulations of the assembly of cluster halos, Gao et al. (2004b)
addressed the latter problem by analysing the rate at which material is added
to the innermost region where the visible galaxy lies.  Their most striking
finding was that while the total mass of the inner $10h^{-1}$kpc
has evolved little since redshift $z\sim 6$, much of the material in the
current core has been added recently from previously distinct objects.  Here
we carry out a similar study of the assembly of the inner cores of our
Millennium Simulation halos.

Following the approach of Gao et al. (2004b), we find the fraction of
the material in the core of each $z=0$ halo which was part of a entirely
different object at each of a series of earlier redshifts. We refer to
this fraction as the ``accreted fraction'' and we then estimate the
distribution of accreted fraction for the halos in each of our four mass
ranges and for accretion since redshifts of 0.5, 1, 2 and 3. For the purposes
of this calculation we define the ``core'' of each $z=0$ halo to consist
of the 100 most bound particles in its main subhalo. Each particle is
considered to be part of a disjoint object at some earlier redshift
(and thus part of the accreted fraction) if at that time it was more than
$100h^{-1}$kpc (physical) from the centre of the largest progenitor of
the core (defined by calculating the mutual gravitational potential
of all 100 core particles and picking the particle with the lowest value).

The results of this exercise are shown in Fig.~\ref{fig:fig5}. Each panel
refers to one of our ranges of halo mass and contains four lines giving the
cumulative distributions of accreted fraction since the four redshift
indicated by the colour code, i.e. the fraction of all halos for which the
accreted fraction exceeds the percentage given in the abscissa.  A label in
each panel also gives the average distance from the centre of the final halo
for the 100 particles used to define the core. This turns out to be close to
$10h^{-1}$kpc, independent of mass.

The results are qualitatively similar for all four mass ranges, but
they show quantitative differences of the kind expected from our
earlier analysis. Just under a quarter of all cluster halos have
accreted at least 10\% of their core mass since $z=0.5$, while over
40\% of them have accreted this much since $z=1$. For Milky Way halos
the corresponding fractions are a tenth since $z=0.5$ and a quarter
since $z=1$. A fifth of all Milky Way halos have accreted at least
some core mass since $z=0.5$ and 40\% of them since $z=1$. It is
interesting that the mass dependence of these curves decreases with
increasing redshift and indeed no significant mass dependence is
detected for the distributions of accreted fraction since $z=2$ or 3.
The conclusion would seem to be that merger-related accretion into the
inner regions where the galaxy resides is predicted to be significant
for a small but non-negligible fraction of isolated halos similar to
that of the Milky Way. As pointed out by Toth \& Ostriker (1992) and
further studied by Velazquez \& White (1999), such late accretion may
observably affect the thickness of the stellar disks of these 
galaxies. The results for cluster halos are consistent with those of
Gao et al. (2004b); many objects have accreted a significant fraction
of the mass in their inner $10h^{-1}$kpc since $z=0.5$.

\section{Discussion}
\begin{figure*} \hspace{-0.5cm}
\resizebox{9cm}{!}{\includegraphics{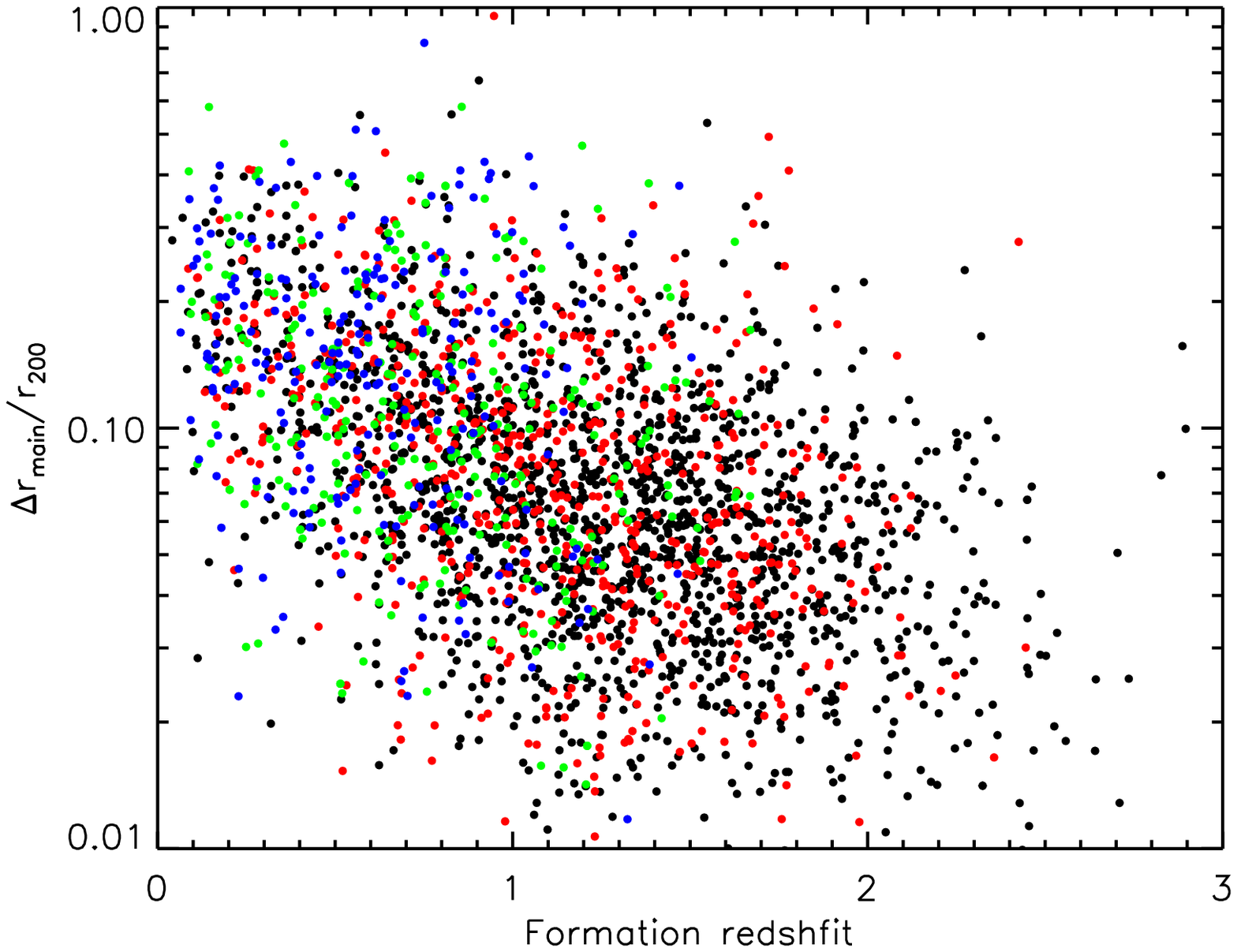}}%
\hspace{-0.1cm}\resizebox{9cm}{!}{\includegraphics{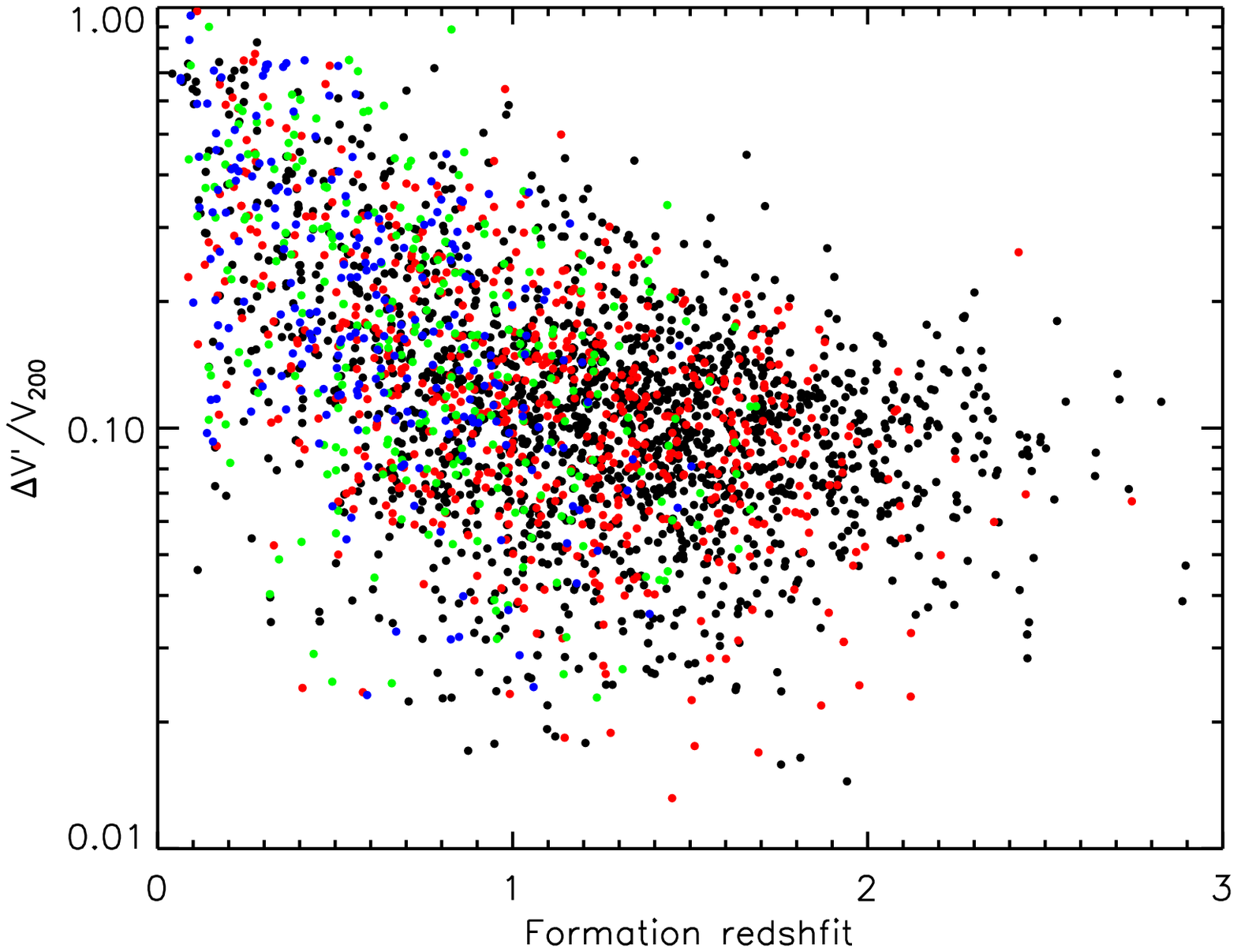}}
\caption{Asymmetries of haloes as a function of their formation time. (Left)
  The spatial offset between the core and the bulk of the main subhalo,
  normalised by $r_{200}$, the radius of the corresponding {\small SO} halo,
  is plotted against halo formation time, defined as the latest epoch at which
  the mass of the dominant progenitor was less than half the final
  mass. (Right) The corresponding velocity offset $\Delta
  V'=|\vec{V}_{\rm core} - \vec{V}_{\rm main}|$,   normalised by
  $V_{200}$, is plotted against the same formation time. Different
  colours in these plots   refer to haloes of different mass: black --
  $[2,4] \times 10^{12}h^{-1}{\rm M_\odot}$; red -- $[0.7,2] \times
  10^{13}h^{-1}{\rm M_\odot}$; green -- $[0.7, 2] \times
  10^{14}h^{-1}{\rm M_\odot}$; blue -- all haloes more massive than $2
  \times 10^{14}h^{-1}{\rm M_\odot}$.}
\label{fig:fig6}
\end{figure*}

In this short paper, we have used the excellent statistics
provided by the very large Millennium Simulation (Springel et al.
2005) to study asymmetries in the inner regions of dark haloes and
their possible relation to the accretion of external material onto
these regions. These asymmetries can be thought of as resulting
from excitations of the oscillation modes of quasi-equilibrium
haloes. They may be related to visible features of galaxies such
as warps, lopsided disks, asymmetric rotation curves, polar rings,
stellar streams, etc. A proper exploration of this relationship
would, of course, require detailed treatment of the visible
components in addition to the dark matter.

For present-day haloes we find the typical amplitude of
asymmetries to depend quite strongly on halo mass. A fifth of all
cluster haloes have density centres offset from their barycentre
by more than 20\% of their virial radius, while only 7\% of Milky
Way haloes have such a large asymmetry. About 40\% of all cluster
haloes have a core velocity which differs from their barycentre
velocity by more than a quarter of the characteristic circular
velocity, whereas only 10\% of Milky Way haloes have such large
velocity offsets. This mass dependence of asymmetries is mirrored,
albeit somewhat more weakly, in the statistics of material
accretion onto the inner halo. About 25\% of all cluster haloes
have acquired at least a quarter of the mass currently in their
inner $10$kpc through mergers since $z=1$. The corresponding
percentage for Milky Way haloes is 15\%.

Our argument that the asymmetries are related to the recent assembly
history of haloes can be demonstrated directly on a halo-by-halo
basis. In Fig.~\ref{fig:fig6}, we plot position (left panel) and
velocity (right panel) asymmetry as a function halo formation time,
which we here define as the time when half of the current halo mass
was first assembled in a single object. This definition dates back to
Lacey \& Cole (1993) and is often used in numerical studies of halo
assembly and clustering (e.g. Gao, Springel \& White 2005). The
different colours in these plots refer to haloes of different
mass. There is a clear anticorrelation between asymmetries of both
types and halo formation time. This is visible not only between haloes
of differing mass, but also among haloes of the same mass. Indeed, the
four different mass groups appear to follow the same relations in
these plots. The correlation is weakest for velocity asymmetries of
low-mass haloes. This may be due to noise in our estimates of the core
velocities of these systems (see Section~3).  Thus it appears that
ongoing accretion events associated with halo assembly continually
excite oscillations of the inner cores which gradually damp between
events. Further work to link this directly with observed kinematic and
photometric distortions of galaxies, as well as with distortions in
X-ray and lensing images of clusters, would clearly be worthwhile.
Published studies of cluster asymmetry show position offsets
comparable to those that we find (Mohr et al. 1995, Lazzati \&
Chincarini 1998, Kolokotronis et al.  2001), while typical differences
between the mean velocities of galaxy clusters and those of their
central galaxies also appear similar to the offsets we predict
(Zabludoff, Huchra \& Geller 1990; van den Bosch et al. 2005b).

\section*{Acknowledgements}
The authors are grateful to the Virgo Consortium, and in particular to
Volker Springel, for the tremendous amount of work needed to carry out
the Millennium Simulation and to make its results available for analysis.
The simulation was carried out on the Regatta supercomputer of the
Computing Center of the Max-Planck-Society in Garching. G.L. also thanks
Yipeng Jing for constructive discussions.
\label{lastpage}

\end{document}